\documentclass[prb,superscriptaddress,twocolumn]{revtex4-1}
\usepackage{bm}
\usepackage{graphicx}
\usepackage{float}
\usepackage{color}
\begin{document}

\title{Hidden type-II Weyl points in the Weyl semimetal NbP}
\author{Shu-Chun Wu} \author{Yan Sun} \author{Claudia Felser} 
\affiliation{Max Planck Institute for Chemical Physics of Solids, D-01187 Dresden, Germany}
\author{Binghai Yan}
\email{binghai.yan@weizmann.ac.il}
\affiliation{Department of Condensed Matter Physics, Weizmann Institute of Science, 7610001 Rehovot, Israel}

\date{\today}
\begin{abstract}
As one of Weyl semimetals discovered recently, NbP exhibits two groups of Weyl points with one group lying inside the $k_z=0$ plane and the other group staying away from this plane. All Weyl points have been assumed to be type-I, for which the Fermi surface shrinks into a point as the Fermi energy crosses the Weyl point. 
In this work, we have revealed that the second group of Weyl points are actually type-II, which are found to be touching points between the electron and hole pockets in the Fermi surface. 
Corresponding Weyl cones are strongly tilted along a line approximately $17^\circ$ off the $k_z$ axis in the $k_x - k_z$ (or $k_y - k_z$) plane, violating the Lorentz symmetry but still giving rise to Fermi arcs on the surface.
Therefore, NbP exhibits both type-I  ($k_z=0$ plane) and type-II ($k_z \neq 0$ plane) Weyl points. 
\end{abstract}

\maketitle

\section{Introduction}

Weyl semimetals (WSMs)~\cite{Wan2011,volovik2003universe,murakami2007,Balents2011,Yan2017,Armitage2017} were recently found in four transition-metal monopnictides, TaAs, TaP, NbAs and NbP, by theoretical predictions~\cite{Weng2015,Huang2015} and angle-resolved photoemission spectroscopy (ARPES)
~\cite{Xu2015TaAs,Lv2015TaAs,Yang2015TaAs,Liu2016NbPTaP,Xu2015NbAs,Belopolski2016NbP,Xu2016TaP,Souma2015NbP}.
In the bulk, these materials exhibit Weyl points through which conduction and valence bands cross each other linearly in the three-dimensional momentum space. 
At the Weyl point, the Fermi surface usually shrinks into a point. 
The quasiparticle excitations near the Weyl point behave effectively as a kind of massless relativistic particles known as Weyl fermions in the standard model.
These Weyl points are monopoles of the Berry curvature field and thus lead to topological boundary states on the surface. 
The topological surface state connects a pair of Weyl points with opposite chiralities, resulting in an unclosed curve in the Fermi surface, called the Fermi arc.
Fermi arcs serve a hallmark for the detection of Weyl fermions by ARPES and scanning tunneling microscopy~\cite{Inoue2016, Batabyal2016, Zheng2016}. 
Soon transport experiments have been extensively studied on these materials for large magnetoresistance (MR)~\cite{Shekhar2015,Ghimire2015NbAs,Huang2015anomaly,Zhang2016ABJ,Wang2015NbP,Luo2015,Moll2015}, the chiral anomaly effect~\cite{Huang2015anomaly,Zhang2016ABJ,Wang2015NbP,Niemann2017}, the gravitational anomaly effect~\cite{Gooth2017}, optical response~\cite{Ma2017,Wu2017}, and even catalysis~\cite{Rajamathi2017}. 

\begin{figure}[htb]
\centering
\includegraphics[width=8cm]{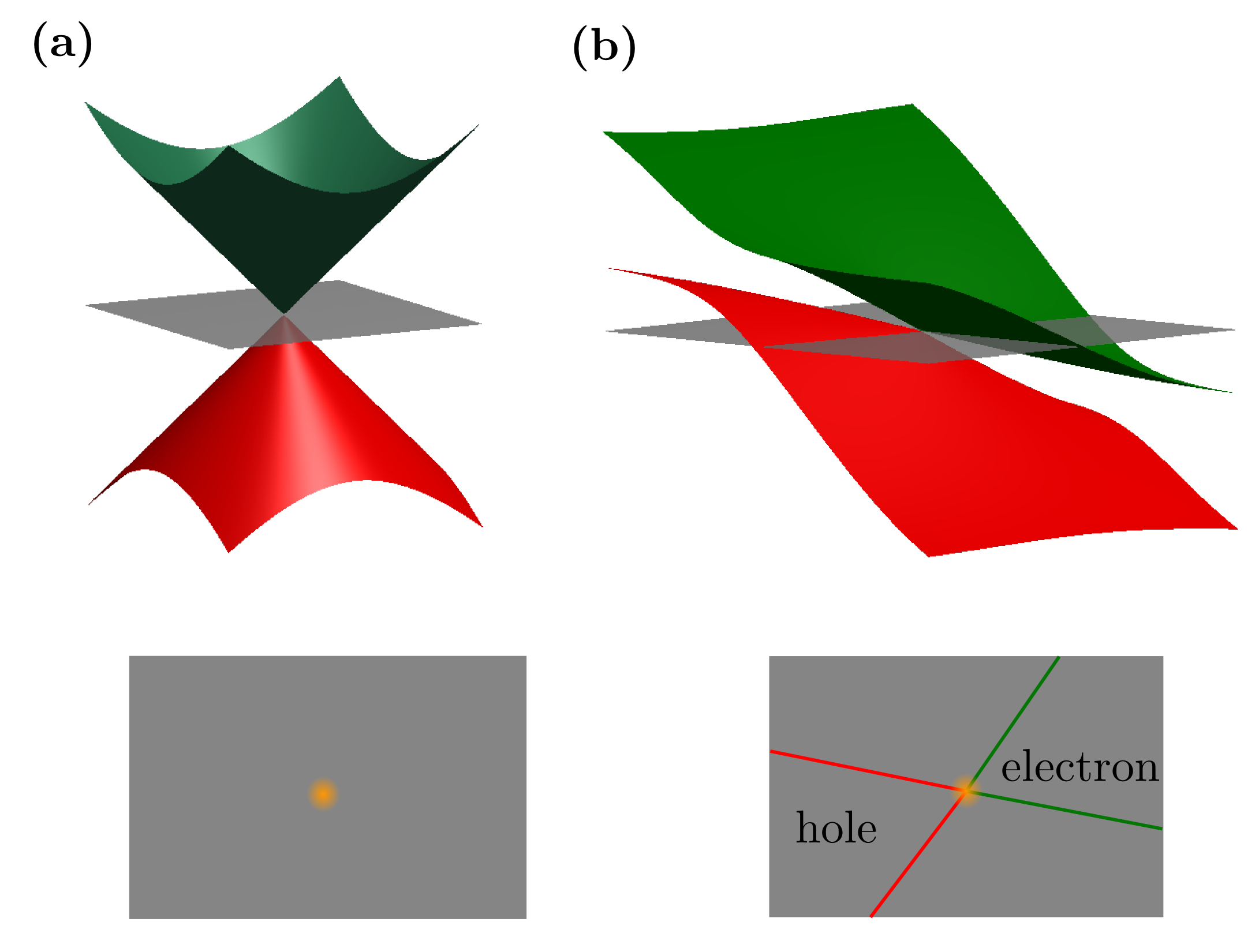}
   \caption{Schematic diagrams of different types of Weyl points. (a) The type-I Weyl cone. When
the Fermi energy is sufficiently close to the Weyl points, the Fermi surface
shrinks to a point. (b) The type-II Weyl cone. When the Fermi energy
crosses the Weyl point, the Weyl point become the touching
point between hole and electron pockets in the Fermi surface, due to the strong tilting of the Weyl cone.}
\label{fig:type}
\end{figure}

Very recently a new type of WSMs were anticipated~\cite{Soluyanov2015WTe2,Xu2015} and discovered in MoTe$_2$~\cite{Sun2015MoTe2,Wang2015MoTe2,Huang2016,Deng2016,Jiang2016,Liang2016}, called the type-II WSM. 
The Weyl cone is strongly tilted so that the Fermi velocity reverses its sign along the tilting direction (see Fig.~\ref{fig:type}).
The type-II WSM exhibit a very different Fermi surface from a normal WSM (referred to as type-I) at the Weyl point.
It has finite Fermi pockets, where the touching point between the electron and hole pockets is the type-II Weyl point. 
Although it also exhibits Fermi arcs, the type-II Weyl cone is expected to demonstrate direction-dependent chiral anomaly effect~\cite{Soluyanov2015WTe2} and large photocurrents~\cite{Chan2017}.

For long, TaAs, TaP, NbAs and NbP are believed to exhibit topologically equivalent band structures as the type-I WSM, where the length of Fermi arcs as well as the Weyl point separation varies as the spin-orbit coupling (SOC) is different~\cite{Sun2015arc,Liu2016NbPTaP}. 
In this work, we report that a group of Weyl points in NbP, which has the weakest SOC among four compounds, are actually type-II by bulk and surface band structure calculations. This fact is overlooked in previous research, mainly because these Weyl points lie above the Fermi energy and cannot be directly accessed in ARPES and transport measurements, and also because the tilting direction of the Weyl cone is away from ordinary crystallographic axes.

\section{Methods}

The band structure calculations were performed with the density-functional theory within  the generalized gradient approximation, which is implemented in the Vienna \textit{ab-initio} simulation package (\textsc{vasp})~\cite{vasp}. The SOC was included. 
We project the $ab~initio$ wave functions to the localized Wannier functions~\cite{mostofi2008wannier90} and constructed the tight binding Hamiltonian for bulk NbP, to interpolate the band structure and Fermi surface in the Brilloui zone by dense $k$-grids.
For the surface states, we used two models. One is a semi-infinite surface model constructed by the tight-binding parameters based on the bulk Wannier orbitals. The other is a slab model with a thickness of 15 bulk unit-cells, calculated in a fully $ab~initio$ way. Two models give different Fermi arc states due to different boundary conditions, but exhibit the same topology in the Fermi surface. 

\section{Results}
\begin{figure}[htb]
\centering
\includegraphics[width=8cm]{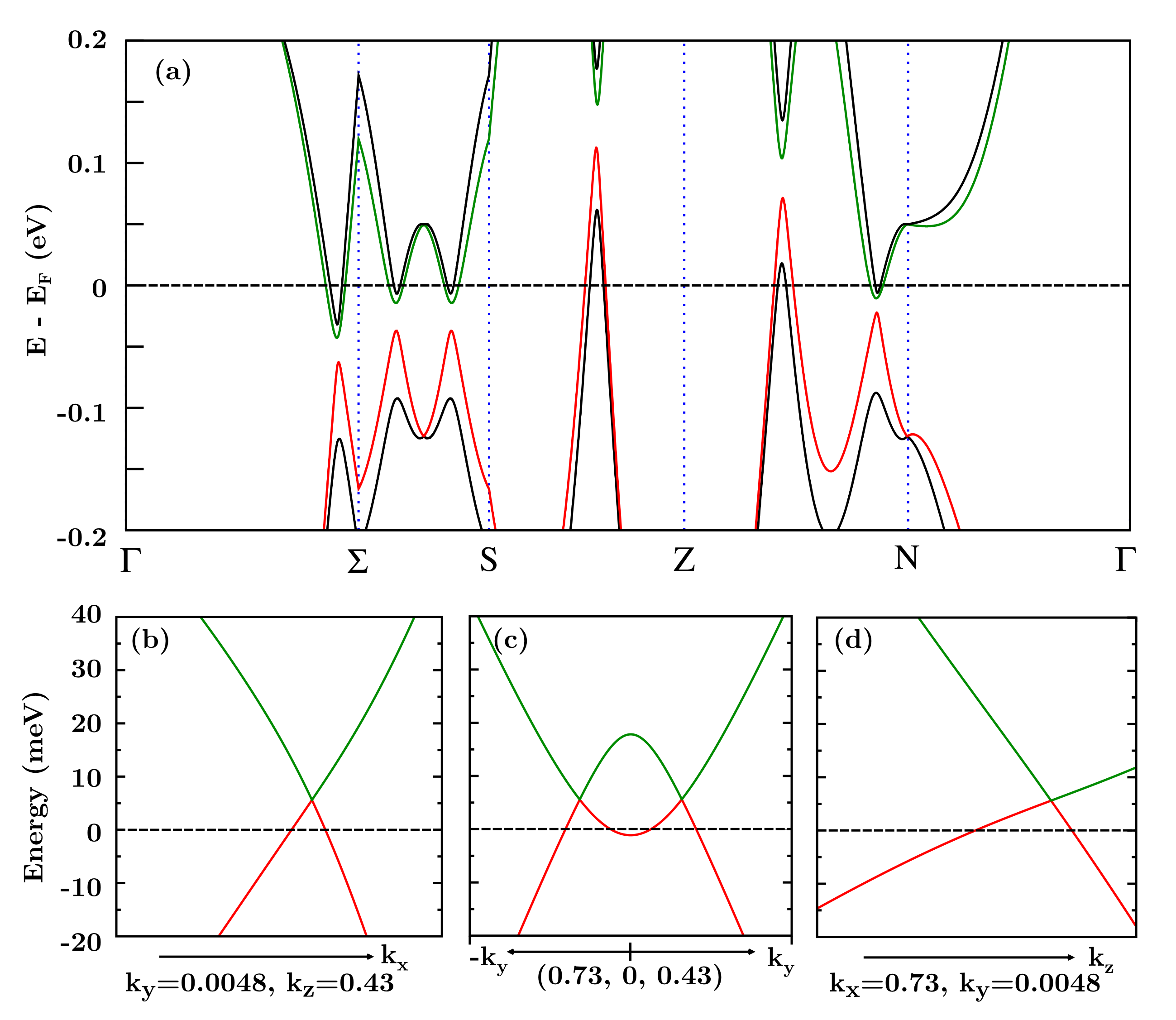}
   \caption{(a) Bulk band structure of NbP. (b)-(d) The band structure along 
(b) k$_x$, (c) k$_y$ and (d) k$_z$ directions. The red line is the highest 
valence band and the green line is the lowest conduction band. k$_x$ and k$_y$ are in unit 2$\pi/a$, and k$_z$ is in unit 2$\pi/c$.}
\label{fig:band}
\end{figure}

The energy dispersion of NbP presents as a semimetallic state with
tiny electron and hole pockets cutting the Fermi level. Due to
inversion symmetry breaking, spin degeneracy splits for all the bands~\cite{Weng2015,Huang2015},
as shown in Fig.~\ref{fig:band}(a). Based on the calculated
band structure we have achieved good agreement with experimental
magnetic quantum-oscillation measurements, which confirmed the
accuracy of our calculations~\cite{Klotz2016, Reis2016, Sergelius2016}. Consistent with
previous reports, there are totally twelve pairs of Weyl points
in the whole Brillouin zone (BZ), with four pairs in the $k_z=0$
plane (W1) and the others in the plane around $k_{z}\sim\pm\pi/c$ (W2),
which are 57 meV bellow and 5 meV above the Fermi level in energy space,
respectively~\cite{Weng2015,Huang2015,Klotz2016, Reis2016, Sergelius2016}. 
The Fermi surface topology revealed that W1-type Weyl points are type-I.~\cite{Klotz2016}. 

The energy dispersion crossing the W2 Weyl point along high symmetry directions of $k_x$, $k_y$ and $k_z$ present very good type-I features, as presented in Fig.~\ref{fig:band}(b-d). That should be a simple reason why W2 Weyl points was assumed to be a type-I. However, it cannot exclude the possibility that the W2 Wely cone is strongly tilted along some low-symmetric direction.

\begin{figure}[htb]
\centering
\includegraphics[width=8cm]{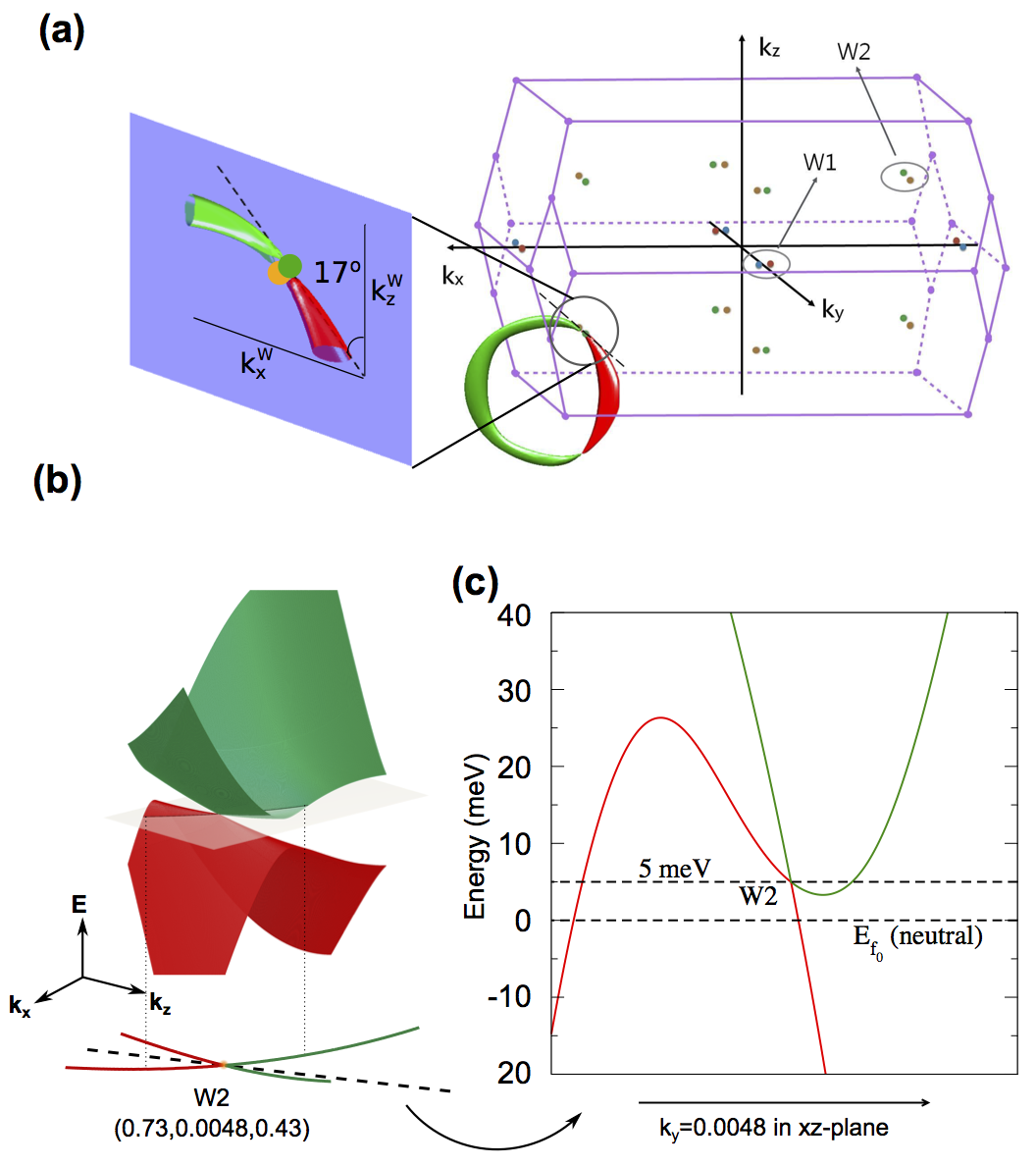}
   \caption{(a) Eight pairs of Weyl points are denoted by green and yellow dots to 
represent opposite chirality of type-II Weyl points, and four pairs of Weyl points 
are denoted by blue and red dots to represent opposite chirality of type-I Weyl points. 
Red and green pockets represent hole and electron pockets, respectively. The black dash 
line indicates the special direction for non-chiral anomaly. (b) The 3D band structure 
of the W2 Weyl cone in the k$_x$-k$_z$ plane is heavily tilted, which shows the type-II 
character of the W2 nodes in NbP. The energy contour at the bottom corresponds to the 
Fermi level at W2 Weyl point. Two bands of upper and lower cones are indicated as green 
and red lines. The tilted line in (a) is also projected into the contour. (c) The band 
structure along the tilted line in (a). The cross-section of the 3D band structure 
shows that the W2 node lies 5~meV above the intrinsic Fermi level $E_{f_0}$(neutral). 
$k$ coordinates are in unit of reciprocal vectors.}
\label{fig:weyl}
\end{figure}

For further identifying the feature of the Weyl point in NbP,
we analyzed the Fermi surface accordingly. As it is already known
from quantum oscillation and previous calculations, when the Fermi
level is 5 meV bellow Weyl points, both of the electron and hole
pockets present as banana shape with strong anisotropy~\cite{Klotz2016}.
The Fermi surfaces can be separated into two branches in the reciprocal
space without any touching point. However, if we shift up the chemical
potential exactly to the energy of W2 Weyl points , it is found that electron and hole pockets touches each other 
 in $k_{z}\sim\pm\pi/c$ planes. Here, W2 Weyl points are exactly the touching points, presenting clearly a type-II feature. 
 Taking the Fermi surfaces with $M_y$ symmetry as the example, Fig.~\ref{fig:weyl}(a), we can see that the
electron and hole pockets linear touching each other along the
direction of 17 $^\circ$ off the $k_z$ axis. So the Weyl cone is strongly
titled in the $k_{y}=k_{y}^{W}$ planes (the position of Weyl point W2
is ($k_{x}^{W},k_{y}^{W},k_{z}^{W}$)). As shown in Fig.~\ref{fig:weyl}(b), The 2D
cross-section of Fermi surfaces in $k_{y}=k_{y}^{W}$ plane
forms a linear crossing but not shrink to a point. We also analysed
the energy dispersion crossing the Weyl point along the titled
direction, Fig.~\ref{fig:weyl}(c), which is completely different from that along
high symmetry lines as given in Fig.~\ref{fig:band}(b-d). 

The type-II Weyl point is further confirmed by the surface state.
Though both type-I and type-II Weyl points can induce the non-closed
surface Fermi arc, the shapes of them are different. In type-I WSMs,
the bulk Fermi surface shrinks to a point at Weyl point. Hence, when
the energy is fixed at the Weyl point, the Fermi arc will terminate
at two isolated points without any bulk density of states. While in
type-II WSMs, one can expect to observe the linear touching of surface
projected electron and hole pockets at the Weyl points, where are
terminations of the Fermi arcs. In general, anion-terminated (P)
surface in (001) direction was usually reported for the as-cleaved
surface in ARPES for NbP~\cite{Liu2016NbPTaP,Belopolski2016NbP,Souma2015NbP}.
Since NbP is not a layered material, the chemical bonding is very
strong, and the charge redistribution plays an important role for
the detailed shape of the surface Fermi arc states. Therefore,
in this work we have first analysed the surface state by a tight-binding semi-infinite model, which provides a clear understanding
of the topological Fermi arc state. Further, we also studied the surface state with surface
charge redistribution taken into account by fully \textit{ab-initio}
slab calculations. 
Tight-binding and \textit{ab-initio} show different surface states, but the same topology.

\begin{figure}[htb]
\centering
\includegraphics[width=8cm]{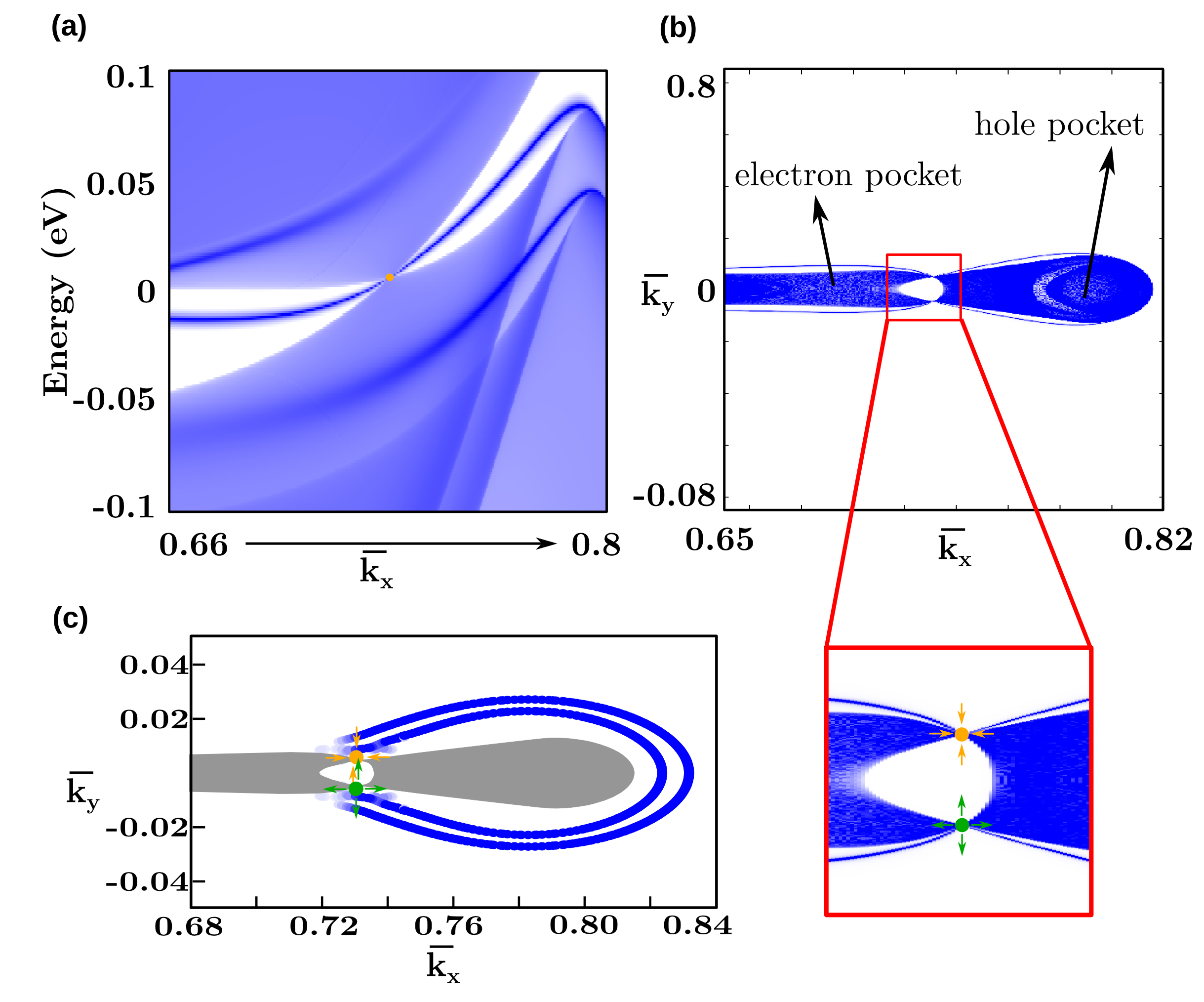}
   \caption{(a) The band structure along $\bar{\mathrm{k}}_x$ ($\bar{\mathrm{k}}_y=0.0048$) 
and (b) the Fermi surface of the semi-infinite surface. (c) The Fermi surface of slab model 
by DFT. The gray region is bulk states. $k_x$ and $k_y$ are in unit 2$\pi/a$.}
\label{fig:Fermi_arc}
\end{figure}

For convenience we just focus on one pair of Fermi arcs near the
$M_y$ mirror plane. The surface band structure with P termination
along $\bar{\mathrm{k}}_x$ with fixed $k_{y}=k_{y}^{W}$ is given in
Figure~\ref{fig:Fermi_arc}(a). Since two pairs of W2 points with the
same chirality are projected to the same point in the (001) 2D BZ,
two Fermi arcs are expected from one pair of projected Weyl points.
From surface energy dispersion in Fig.~\ref{fig:Fermi_arc}(a) one can easily see that
the conduction bands and valence bands linearly touching each other
at the projected Weyl point, with two surface bands crossing this
points. Since no other surface band appears, we expect that these
two states are just the Fermi arcs related states. For further
understanding, we analyzed the projected surface state with chemical
potential lying at the Weyl point, as shown in Fig.~\ref{fig:Fermi_arc}(b). One pair
of projected Weyl points with opposite chirality are presented as
the linear touching point of electron and hole pockets, and two clear
non-closed Fermi arcs originated from the linear touching points.
Therefore, the type-II Weyl point are directly confirmed by the
co-existence of linear touching of the projected bulk Fermi surfaces
and corresponding Fermi arcs.

Though the half-infinite tight binding model provides the correct
understanding for the topological surface Fermi arc, the calculated
surface state does not consistent with experiment ARPES measurement
due to the lack of inclusion of surface charge
redistribution.
Therefore, in order to stimulate the realistic surface states, we also
employed \textit{ab-initio} calculations with a thick slab model. Though two kinds of
calculations give the same topology, where one close loop with one
projected Weyl point inside crosses surface FSs twice, the details
of Fermi arc states are very different. In the former calculations,
the two Fermi arcs extended in two opposite directions along $k_x$,
whereas two surface state just locate at the same side of Weyl points
in the fully \textit{ab-initio} calculations, see Fig.~\ref{fig:Fermi_arc}(c). In the slab model,
 the lengths of surface Fermi arcs increases and stay away from the bulk pockets.
Therefore, if one can dope the sample with more electrons, it is
expected to observe the linear touching between projected electron
and hole pockets at Weyl points, with surface Fermi arcs terminated
at these two linear crossing points.

\section{Summary}

In conclusion, we have predicted the existence of type-II Weyl
fermions in NbP based on electronic band structure calculcations.
We revealed that the Weyl cone is tilted strongly along a specific direction, breaking the Lorentz symmetry.
 Since the tilting direction is away from ordinary $a$ and $c$ axes, one may still expect the chiral anomaly effect
along $a$ or $c$ directions. Since the type-II Weyl point is only
5 meV above the Fermi level, weakly electron doping may lead to the
observation of the type-II Weyl points by ARPES. 

\begin{acknowledgments}
This work was financially supported by the ERC (Advanced Grant No. 291472 "Idea Heusler").
B.Y. acknowledges the financial support of the Ruth and Herman Albert Scholars Program for New Scientists in Weizmann Institute of Science and the German-Israeli Foundation (GIF Grant no. I-1364-303.7/2016).
\end{acknowledgments}

%

\end{document}